\documentclass{aastex}

\def\v1n{{\cal U}^N_1}

\def\sss{\scriptscriptstyle}
\def\phih2h2{\phi_{\sss {\rm H_2-H_2}}}
\def\phh2{\phi_{\sss {\rm H-H_2}}}

\def\Teff{T_{\rm eff}}
\def\sqr#1#2{{\vcenter{\vbox{\hrule height.#2pt
  \hbox{\vrule width.#2pt height#1pt \kern#1pt
  \vrule width.#2pt}
  \hrule height.#2pt}}}}

\def\wig#1{\mathrel{\hbox{\hbox to 0pt{%
          \lower.5ex\hbox{$\sim$}\hss}\raise.4ex\hbox{$#1$}}}}

\slugcomment{to be published in The Astrophysical Journal}
\shortauthors{Geballe et al.}
\shorttitle{Observations and Modeling of Gl~570D}

\begin{document}

\title{Infrared Observations and Modeling of \\one of the Coolest T
Dwarfs, Gl~570D}

\author{T. R. Geballe\altaffilmark{1}, D. Saumon\altaffilmark{2}, S. K.
Leggett\altaffilmark{3}, G. R. Knapp\altaffilmark{4}, M. S.
Marley\altaffilmark{5}, and K. Lodders\altaffilmark{6}}

\altaffiltext{1}{Gemini Observatory, 670 N. A'ohoku Place, Hilo, HI
96720; tgeballe@gemini.edu}

\altaffiltext{2}{Department of Physics and Astronomy, Vanderbilt
University, Nashville, TN 37235; dsaumon@cactus.phy.vanderbilt.edu}

\altaffiltext{3}{Joint Astronomy Centre, 660 N. A'ohoku Place, Hilo, HI
96720; s.leggett@jach.hawaii.edu}

\altaffiltext{4}{Princeton University Observatory, Peyton Hall,
Princeton, NJ 08544; gk@astro.princeton.edu}

\altaffiltext{5}{NASA Ames Research Center, Moffett Field, CA 94035 and
Department of Astronomy, New Mexico State University, Las Cruces, NM
88003; mmarley@mail.arc.nasa.gov}

\altaffiltext{6}{Department of Earth and Planetary Science, Washington
University, St. Louis, MO 63130; lodders@levee.wustl.edu}

\begin{abstract}

We have obtained a good quality R$\sim$400 0.8--2.5~$\mu$m spectrum and
accurate photometry of Gl~570D, one of the coolest and least luminous
brown dwarfs currently known. The spectrum shows that Gl~570D has deeper
absorptions in the strong water and methane bands at 1.12--1.17~$\mu$m,
1.33--1.45~$\mu$m, 1.62--1.88~$\mu$m, and 2.20--2.45~$\mu$m and is both
bluer at $J-K$ and redder at $K-L^{\prime}$ than previously observed T
dwarfs. Data analysis using model spectra coupled with knowledge of the
well-understood primary implies that for the same surface gravity,
Gl~570D is about 160$\,$K cooler than Gl 229B. For an age range of
2--5~Gyr Gl~570D has an effective temperature in the range 784--824~K, a
log gravity in the range 5.00--5.27~cm~s$^{-2}$, and a luminosity in the
range 2.88-2.98~$\times$~10$^{-6}$~L$_\odot$.

\end{abstract}


\section{Introduction}

Brown dwarfs with effective temperatures of 1800--3000~K have spectra
closely resembling those of red dwarf stars at similar temperatures, and
the two types of objects share the same classification scheme, either as
M-types in roughly the upper two-thirds of this range or as early L-types
in the lower one-third. The later L-types (with T$_{eff}$ below about
1800~K) are solely the domain of brown dwarfs, but the visible and short
wavelength IR spectra of these objects are a steady continuation of the
trends seen in the spectra of early L-types, with alkali metal lines,
metal hydride bands, and water vapor bands strengthening with decreasing
temperature. As the surface of a brown dwarf cools below approximately
1300--1500~K, however, an important chemical change occurs as methane
(CH$_{4}$) attains a large abundance at the expense of carbon monoxide
(CO). The many strong infrared absorption bands of CH$_{4}$ and the
disappearance of condensates (which significantly affect the spectra of
L-types) below the photosphere substantially alter the spectral appearance
of the brown dwarf, the former making it totally unlike any stellar
object. Once this happens it is thought that dramatic changes in the short
wavelength infrared spectrum do not occur again until water vapor begins
to condense, at temperatures of about 400~K.

Nakajima et al. (1995) and Oppenheimer et al. (1995) used near-infrared
imaging and spectroscopy to discover Gliese~229B, the first brown dwarf
for which absorption bands of methane and water dominate the 1-2.5~$\mu$m
region. Since early 1999 roughly a dozen additional brown dwarfs with
1--2.5~$\mu$m spectra similar to Gl~229B have been reported (e.g., Strauss
et al. 1999; Burgasser et al. 1999; Cuby et al. 1999; Tsvetanov et al.
2000; Burgasser, Kirkpatrick, \& Brown 2001), largely by the Sloan Digital
Sky Survey (SDSS) and the Two Micron All Sky Survey (2MASS), and are
currently known as T dwarfs. The similarity of these spectra attests to
the expected slow evolution of the short wavelength infrared spectrum of a
T dwarf with decreasing temperature, once methane has become the dominant
reservoir of carbon. Recently dwarfs in the L-T transition region near
T$_{eff}~\sim$~1500~K, with detectable CO and CH$_{4}$ in the
1.5-2.5~$\mu$m region, also have been identified (Leggett et al. 2000;
Geballe et al., in preparation) and absorption by the fundamental band of
CH$_{4}$ near 3.3~$\mu$m has been found in mid to late L-type dwarfs by
Noll et al. (2000).

The discovery of brown dwarfs with temperatures much lower than Gl~229B is
made difficult by their lower luminosities and the eventual shift with
lower temperatures of the bulk of their emissions toward longer
wavelengths, where ground-based observations are more difficult. However, 
a few brown dwarfs found by 2MASS have temperatures and luminosities
significantly below those of Gl~229B - type dwarfs. One of these is
Gliese~570D, discovered by Burgasser et al. (2000) and estimated by them
to have an effective temperature of 750$\pm$50~K and luminosity of
(2.8$\pm$0.3)$\times$10$^{-6}$~L$_{\odot}$.  We have obtained a medium
resolution 0.8--2.5~$\mu$m spectrum of this object, which represents a
substantial increase in resolving power, wavelength coverage, and
signal-to noise ratio over that reported by Burgasser et al. (2000).  
Here we present this spectrum and new photometry, along with a detailed
analysis which yields improved values for temperature, luminosity and
surface gravity.

\section{Observations and data reduction}

\subsection{Photometry}

JHK infrared photometry was obtained for Gl~570D on UT 2000 February 1
with the UFTI camera on the 3.8~m United Kingdom Infrared Telescope
(UKIRT) on Mauna Kea.  L$^{\prime}$ photometry was obtained on 2000 May 15
with IRCAM, also on UKIRT.  Weather conditions were photometric on both
nights. No color term is needed to convert the L$^{\prime}$ data to the
established ``UKIRT'' L$^{\prime}$ system; however, the J, H and K
photometric systems are different from the ``UKIRT'' system due to
significant differences in the filter set.  The new photometric system is
referred to as the ``MKO--NIR'' system, as the recently specified Mauna
Kea--Near Infrared filter set defines the system.  Transformations between
the established UKIRT system and the new MKO--NIR system are presented in
Hawarden et al. (2000).

\subsection{Spectroscopy}

Spectra of Gl~570D were acquired at UKIRT on UT 2000 March 13-15,
utilizing the facility 1-5$\mu$m 2D array spectrometer, CGS4 (Mountain et
al. 1990). An observing log is provided in Table~1. Sky conditions were
excellent. The observations employed CGS4's 40~lines/mm grating in second
order shortward of 1.4~$\mu$m and in first order longward of that
wavelength, and a 2-pixel (1.2$\arcsec$) wide, 80$\arcsec$ long slit.
Observations were obtained in the stare: nod-along-slit mode, with a nod
of 7.32$\arcsec$ (12 rows on the array). Prior to observing in each
waveband the spectrum of the nearby F5V star, HD~126819 was measured for
the purpose of flux calibration and removal of telluric absorption lines.  
Spectra of argon and krypton arc lamps were obtained for wavelength
calibration.

\begin{deluxetable}{cccc}
\tablecolumns{4}
\tablecaption{Spectroscopy of Gl 570D}
\tablewidth{0pt}
\tablehead{
\multicolumn{1}{c}{UT Date} &
\multicolumn{1}{c}{Wavelength Range} &
\multicolumn{1}{c}{Resolution} &
\multicolumn{1}{c}{Integration time} \\
\multicolumn{1}{c}{} &
\multicolumn{1}{c}{$\mu$m} &
\multicolumn{1}{c}{$\mu$m} &
\multicolumn{1}{c}{seconds} 
}
\startdata
2000 March 13 & 1.88--2.52 & 0.0050 & 2400 \\
2000 March 13 & 1.43--2.07 & 0.0050 & 1920 \\
2000 March 14 & 1.03--1.35 & 0.0025 & 2880 \\
2000 March 15 & 0.79--1.11 & 0.0025 & 4800   
\enddata
\end{deluxetable}

The data reduction employed Figaro routines to extract source spectra
from the spectral images produced by the CGS4 array, wavelength-calibrate
them, ratio them, and perform an initial flux calibration. It was found
that where the spectral segments overlapped or adjoined one another (at
1.0--1.1~$\mu$m, 1.3--1.5$\mu$m, and 1.9--2.1~$\mu$m) they did not match
perfectly; typical disagreements were 10 percent. The mismatches are
probably due to variations in guiding accuracy and in seeing during the
observations, as well as probable inaccuracies in the IR magnitudes of
the calibration star used in the reduction. The spectral segments were
scaled by small factors in order to adjoin to one another smoothly and to
match the photometry as well as possible, which they do to $\pm$5\%.

\section{Results}

Table 2 gives the JHKL$^{\prime}$ values in the ``MKO--NIR'' system as
well as in the ``UKIRT'' system. The latter has been calculated by
convolving the flux-calibrated energy distribution with the old filter
profiles. These show Gl~570D to be considerably bluer at $J-K$ than
reported by Burgasser et al. (2000) and indeed bluer than all but one T
dwarf reported to date. For example, for the T dwarfs Gl~229B, SDSS~1346,
and SDSS~1624, $J-K$ is in the range $-0.4~\le~J-K~\le~-0.1$ (Leggett et
al. 1999, 2000) whereas for Gl~570D $J-K$ is -0.51$\pm$0.07 (both in the
UKIRT system). This increasing blueness continues the trend that has been
found from the latest L-types through the earliest T-types and on to the
classical T dwarfs such as Gl~229B (Leggett et al. 2000). Furthermore
Stephens et al. (2001) and Marley et al. (2001) find that $K-L'$
increases monotonically through the L and T sequence.  Stephens et al.  
find SDSS 1624 to have the reddest $K-L'$ of five T dwarfs observed at
$2.54\pm0.08$, while we find Gl~570D has $K-L' =3.54\pm 0.05$, again
suggesting that Gl~570D is quite cool.
 
\begin{deluxetable}{rrrrrrr} 
\tablecolumns{7}
\tablewidth{0pt}
\tablecaption{New Photometry of Gl 570D}
\tablehead{
\multicolumn{4}{c}{MKO--NIR} &
\multicolumn{3}{c}{UKIRT} \\
\colhead{J (error)} &
\colhead{H (error)} &
\colhead{K (error)} &
\colhead{L$^{\prime}$ (error)} &
\colhead{J } &
\colhead{H } &
\colhead{K }}
\startdata
14.82 (0.05) & 15.28 (0.05) & 15.52 (0.05) & 12.98 (0.05) & 15.12 & 15.32
& 15.63 \\
\enddata
\end{deluxetable}

The 0.83--2.52~$\mu$m spectrum of Gl~570D (continuous apart from a gap at
1.36--1.44~$\mu$m and slightly smoothed) is shown in Fig.~1. The figure
also contains a spectrum of Gl~229B (Geballe et al. 1996, flux-calibrated
by Leggett et al. 1999) for comparison.  The only atomic feature observed
in the spectrum of Gl~570D is the potassium doublet near 1.25~$\mu$m,
which is considerably weaker than the doublet in Gl~229B and is only just
discernable. The strong absorption bands of CH$_{4}$ and H$_{2}$O define
the overall shapes of the two spectra, which are quite similar. However,
the bands are deeper and in some cases broader in Gl~570D than Gl~229B.
The increasing blueness at $J-K$ with decreasing temperature is thus seen
to be a consequence of the relative degrees of strengthening of the
(already very strong) absorption bands of water and methane. It also is
due in part to the increased opacities of the broad (and blended)
H$_{2}$ pressure-induced 1-0 S(1), S(0) and Q(1) absorptions at
2.12~$\mu$m, 2.22~$\mu$m, and 2.41~$\mu$m (e.g., Borysow, J$\phi$rgensen,
\&
Zheng 1996).

\section{Modeling of Gl 570D}

The following analysis is based on the brown dwarf atmospheric models
described in Burrows et al. (1997) and Saumon et al. (2000) and on the
0.8-2.5~$\mu$m spectroscopy and photometry presented here. Briefly, the
atmospheres are in radiative/convective equilibrium and the molecular
opacities are treated with the k-coefficient method (Goody et al. 1989;
Lacis \& Oinas 1991). Chemical equilibrium is treated as in Burrows et al.
(1997). Gas-phase opacities include Rayleigh scattering, the
collision-induced opacity of H$_2$ and the molecular opacities of H$_2$O,
CH$_4$, NH$_3$, H$_2$S, PH$_3$, CO, VO, and FeH as well as the continuum
opacities of H$^-$ and H$_2^-$. The present models also include the line
opacities of Na, K, and Cs as described in Burrows, Marley, \& Sharp
(2000). It appears that condensates do not play a significant role in
shaping the spectrum of the cooler T dwarfs (Burrows et al. 2000; however,
see Tsuji, Ohnaka, \& Aoki 1999). The opacities of condensation clouds are
not included in the present models, although the chemical equilibrium
calculation takes into account the formation of condensates.  
Particulates are assumed to fall below the spectroscopically accessible
region of the atmosphere. The equation of radiative transfer is solved
including scattering, following Toon (1989).

Compared to the T dwarf Gl~229B, Gl~570D offers the advantage that the
primary star of the system is fairly well studied.  The metallicity of
Gl~570A was determined by Hearnshaw (1976) and Feltzing \& Gustafsson
(1998) to be [Fe/H]=0.01 and $0.00 \pm 0.12$, respectively. We adopt the
solar abundances of Anders \& Grevesse (1989) for modeling Gl~570D.

Using the same monochromatic opacities used to compute the k-coefficients,
high-resolution synthetic spectra were generated from the atmospheric
structures by solving the radiative transfer equation with the Feautrier
method (Mihalas 1978) on a frequency grid with $\Delta \nu=1\,$cm$^{-1}$.
The synthetic spectra were convolved with a Gaussian filter to match the
resolution of the observations.

\subsection{Luminosity, effective temperature and gravity}

Because the present spectrum samples more than half of the flux emitted
by Gl~570D, and because the distance is known to be 5.91$\pm$0.06~pc
(Perryman et al. 1997), we can tightly constrain the possible solutions
for the surface parameters of Gl~570D by using both synthetic spectra and
evolution calculations.  We use a method similar to the one applied to
Gl~229B by Saumon et al. (2000). The solar-metallicity evolution sequence
for cooling brown dwarfs of Burrows et al. (1997) specifies uniquely the
bolometric luminosity $L_{\rm bol}$ of a brown dwarf in terms of the
effective temperature $\Teff$ and the surface gravity $g$. This defines a
surface $L_{\rm bol}(\Teff,g)$.

\begin{deluxetable}{ccccccc}
\tablecolumns{7}
\tablewidth{0pt}
\tablecaption{Physical parameters as a function of surface gravity}     
\tablehead{
\colhead{Model} & \colhead{$\log g$}  &  \colhead{$T_{\rm eff}$}  &    
\colhead{$\log L/L_\odot$}
 & \colhead{Mass} & \colhead{Radius} & \colhead{Age} \\
\colhead{}  & \colhead{(cgs)}  & \colhead{(K)} &   \colhead{}  &     
\colhead{$(M_J)$}  & \colhead{$(R_\odot)$}  & \colhead{(Gyr)}}
\startdata
 A & 4.5 &  734 & $-5.503$ & 15 & 0.1095 & \phs 0.4 \\
 B & 5.0 &  784 & $-5.526$ & 34 & 0.0935 &\phs 2 \phn   \\
 C & 5.5 &  854 & $-5.551$ & 72 & 0.0776 &10 \phn   \\
\enddata
\end{deluxetable}

Our spectrum samples the energy distribution of Gl 570D from 0.83 to
2.52$\,\mu$m.  The integral of the observed flux over wavelength, $L_{\rm
s}=1.581 \times 10^{-15}\,$W/m$^2$, allows us to estimate $L_{\rm bol}$.  
We obtain the bolometric correction factor from our synthetic spectra by
taking the ratio of the flux integrated over the observed wavelength range
to the total emergent flux ($\sigma\Teff^4$).  This synthetic correction
is a function of the spectrum parameters, $\Teff$ and $g$, and is found to
vary between 1.6 and 1.9 in the parameter range of interest here. By
applying the bolometric correction to the empirically-determined flux, we
obtain another relation for the bolometric luminosity, $L_{\rm
bol}(\Teff,g)$. The intersection of the two $L_{\rm bol}$ surfaces gives a
curve $\Teff(g)$ which represents the allowed parameters for Gl 570D.  
For a given gravity, this curve and the cooling sequence give $\Teff$,
$L_{\rm bol}$, the mass, the radius, and the age of the brown dwarf.  The
results are given in Table 3 for three representative gravities.

There are three sources of uncertainty in our determination of $L_{\rm
bol}$.  The distance to Gl~570D is known to a precision of $\pm 1$\% and
the absolute calibration of our spectrum has an uncertainty of $\pm 5$\%.
Finally, the synthetic bolometric correction depends on the reliability of
our synthetic spectra. As discussed below (Section 4.2), there are
systematic differences between the synthetic and observed spectra of T
dwarfs.  Fortunately, these tend to cancel out when calculating integrated
fluxes for the bolometric correction. For example, varying the abundance
of the main absorber, H$_2$O, by 0.1 dex affects the bolometric correction
by only 2\%.  A consistency check can be made with the L$^{\prime}$
measurement, which we did not include in the computation of the bolometric
correction.  The synthetic $L^{\prime}$ magnitude is 14.32, or 0.2 mag
fainter than the observed value.  However, only 7.6\% of the total flux of
Gl~570D is in the L$^{\prime}$ band so the resulting error in the
bolometric correction is $\sim 1.5$\%.  Other modifications of the
synthetic spectrum calculations all indicate that the bolometric
correction is correct to within $\pm 2$\%. The uncertainties in $L_{\rm
bol}$ and $\Teff$ are thus $\pm 6$\% ($\pm 0.03$ dex) and $\pm 12\,$K,
respectively.

Figure 2 shows the models listed in Table 3 superimposed on brown dwarf
cooling curves.  The nearly vertical band running through the center is a
constant luminosity curve with the luminosity of model B ($\log
L/L_\odot=-5.526 \pm 0.03$).  Since all three models A, B, and C have
nearly the same luminosity (see Table 3), models A and C also fall within
that band.  We find that for a given, assumed gravity, Gl~570D is about
160$\,$K cooler than Gl~229B.

\subsubsection{The age of the Gl 570 system}

So far, we have greatly reduced the allowed space for the physical
parameters of Gl~570D (Fig. 2).  However, while $\Teff$ is between 700 and
870$\,$K, and the well-determined luminosity can range from
$\log~L/L_\odot=-5.55$ to $-5.50$, the gravity remains largely
unconstrained, and, with it, the mass and age of Gl~570D. We can reduce
the possible range of gravities by determining the age of the system.  
Fortunately, the primary star, Gl~570A, is a well-observed K4 V star to
which we can apply several age indicators. The Gl~570 system contains a
spectroscopic binary, Gl 570BC. Based on the projected separation of the
A--BC system ($147\,$AU), the mass of the primary ($\sim~0.7~M_{\odot}$),
and the mass of the BC pair ($0.976~M_\odot$), the orbital period of the
A--BC system is of the order of 1400 years (Forveille et al. 1999).  By
comparison, the rotation period of Gl~570A is only 40 days (Cumming,
Marcy, \& Butler 1999).  We can safely assume that there has been no
spin-up of the primary star due to tidal interaction with the BC pair and
that we can use stellar activity indicators as if Gl~570A were an isolated
star.

The X-ray luminosity of Gl~570A is $\log L_X {\rm (erg/s)}=27.72$ (Schmitt
\& Kahabka 1995) or $\sim 5$ times lower than Hyads of the same spectral
type (Stern, Schmitt, \& Kahabka 1995), clearly indicating that Gl~570A is
older than the Hyades cluster. Gl~570A has old disk kinematics (Leggett
1992), and thus is older than $\sim 1.5\,$Gyr (Eggen 1989). The BC pair
has a combined spectral type of M1 V, with individual masses of
$0.586~M_\odot$ and $0.390~M_\odot$, respectively (Forveille et al. 1999).  
The lack of H$_\alpha$ emission from the BC pair indicates that it is
older than $\sim 2\,$Gyr (Soderblom, Duncan, \& Johnson 1991; Hawley,
Gizis, \& Reid 1996).  Thus these indicators suggest a lower bound of
$\sim 2\,$Gyr for the age of the Gl 570 system.

On the other hand, Gl~570A is chromospherically active with a Ca II H and
K emission index of $\log R_{\rm HK}^\prime=-4.49$ (Henry et al. 1996).
This is 0.15 less than Hyads of the same spectral type, indicating that
Gl~570A is older than the Hyades, in complete agreement with our lower
bound for the age.  For comparison, the Sun varies from $\log R_{\rm
HK}^\prime=-5.10$ to $-4.75$ during the solar cycle and is considered
inactive.  Using the age-emission index relation of Donahue (1993; given
in Henry et al. 1996), we obtain an age of 0.8$\,$Gyr for Gl~570A.  
However, if we assume that Gl~570A was observed during a maximum phase of
activity and that its emission index varies with the same amplitude as the
Sun's, then the emission index of Gl~570A might be as low as $\log R_{\rm
HK}^\prime=-4.84$, with a corresponding age of 3.1$\,$Gyr.  Figure 10 of
Soderblom et al. (1991) shows the data used to derive a similar
age-emission index relation.  If we consider the stars showing the most
extreme scatter toward older ages for a given value of $\log R_{\rm
HK}^\prime$, {\it including the error bars on the age determinations}, we
find that to be older than the Sun, Gl~570A would have to have an emission
index of less than -4.95, which is extremely unlikely. We conclude that
the chromospheric activity of Gl~570A indicates relative youth and a very
conservative upper bound of 5$\,$Gyr for its age.

\subsubsection{Optimal parameters}

Figure 2 shows isochrones for our adopted age range for the Gl~570 system.  
The allowed range of gravities for Gl~570D is now considerably reduced to
5.00~$\le$~$\log~g~({\rm cm s^{-2}})~\le$~5.27.  These extremes correspond
to $\Teff\,({\rm K})=784$ (824), $\log L/L_\odot = -5.53$ ($-5.54$), and
$M/M_J=34$ (52), respectively.  Our values for the physical parameters of
Gl~570D agree very well with the estimates of Burgasser et al. (2000).  
The midpoints of these ranges are $\log g=5.13$, $\Teff=804 \pm 12\,$K,
and $L=2.93~\pm~0.18~\times~10^{-6}\,L_\odot$.  The last is less than half
the luminosity of Gl~229B. A meaningful comparison of the masses of
Gl~570D and Gl~229B is not possible since the mass of the latter is poorly
constrained.

We derived the physical parameters simply from the observed integrated
flux, the distance to Gl~570D, synthetic bolometric corrections, and
cooling tracks for brown dwarfs. As for Gl~229B (Saumon et al. 2000), it
was not necessary to fit the observed spectrum with models to derive
$\Teff$ and the gravity. While the synthetic spectra of T dwarfs are
fairly successful at reproducing the observations, there are known
problems which affect a few spectral regions (see below). Fitting the
spectrum over a wide range of wavelengths remains an imprecise exercise
and our method is far more reliable.  By relying only on
wavelength-integrated fluxes, most of the errors in the synthetic spectrum
cancel one another.  Our procedure also uses cooling calculations for the
$L(\Teff,g)$ relation, which is rather insensitive to the details of the
atmospheric surface boundary condition.  Further improvements in the
modeling of brown dwarf atmospheres, principally by the inclusion of
condensation clouds, will have a very modest effect on this relation.

\subsection{Synthetic spectrum}

We have computed synthetic spectra for the models in Table~3. The flux
received at Earth is simply obtained from the radius of each model and the
distance to Gl~570D. The resulting spectra are compared to our
0.83--2.52$\,\mu$m spectrum in Fig.~3.  We emphasize that no fitting of
the spectra has been performed, although spectrum B is a generally better
fit to the observations than either A or C. The overall agreement of the
model with the data is remarkable. This demonstrates both the power and
the internal consistency of our method of analysis and that the overall
spectral energy distribution is well-modeled. Nevertheless, all three
models show particular deviations from the observed spectrum, most of
which are also seen in models for Gl~229B (Allard et al. 1996; Marley et
al. 1996; Saumon et al. 2000).

As discussed by Saumon et al. (2000), the CH$_4$ opacity line list
becomes increasingly incomplete at temperatures above 300$\,$K.  Since
Gl~570D is about 160$\,$K cooler than Gl~229B, we expect to obtain a
better agreement with the observed spectrum in the CH$_4$ bands.  
Indeed, the 2.3$\,\mu$m CH$_4$ band, which is formed at a level where $T
\sim 600\,$K, matches the data extremely well. The 1.6$\,\mu$m band is
formed at $T~\sim~800\,$K, and its overall strength also matches well.
The edge of the band appears at longer wavelengths in the model and the
features in the 1.63--1.72$\,\mu$m region have a much larger amplitude
than in the data, however. This also occurs in models of Gl~229B and
shows the limitation of the CH$_4$ opacity data base.

The H$_2$O bands in the model are too strong, notably in the
1.1--1.2$\,\mu$m, 1.3--1.6$\,\mu$m, and 1.8--2.1$\,\mu$m regions.  This
problem also has plagued cloudless models of Gl~229B and cannot be
resolved by simply invoking a lower abundance of H$_2$O.  Its persistence
suggests that condensates are present in the atmosphere (Tsuji et al.
1999; Ackerman \& Marley, 2001) and that they have a small but detectable
effect on the spectrum.

The heavily pressure-broadened K I resonance doublet at 0.77$\,\mu$m is
mostly responsible for the very weak flux at 0.8--1.0$\,\mu$m (Tsuji et
al. 1999; Burrows et al. 2000). In this model, the opacity of the doublet
is significant up to 1.1$\,\mu$m, yet the flux is overestimated on the
blue side of the 1.08~$\mu$m peak.  Most of this mismatch is due to an
underabundance of neutral K in the cooler part of the atmosphere.

The potassium abundance at levels where $T \wig< 1400\,$K depends on the
overall treatment of condensates and can be modeled in two ways.  One is
to consider that condensates form and remain in local equilibrium with the
gas.  This means that existing high temperature (primary) condensates can
react with the gas at lower temperatures and secondary condensates form
via gas-solid reactions in the upper, cooler part of the atmosphere. This
is the case described as ``no rainout'' by Burrows \& Sharp (1999) and
Burrows et al. (2000) and is the one used in the present analysis.  
However, this scenario for condensate formation may be relevant only for
low-gravity environments such as the solar nebula and stellar outflows and
thus may be unlikely for brown dwarf atmospheres.

A second approach to condensate formation, more applicable for low-mass
stars and brown dwarfs, is the ``condensation cloud formation'' model,
also described as ``rainout'' by Burrows \& Sharp (1999) and Burrows et
al. (2000).  Primary condensates form directly from the gas and are
sequestered in a condensate cloud of finite vertical extent because of the
gravity field.  This prevents the formation of secondary condensates by
gas-solid reactions at lower temperatures (above the cloud) that deplete
the abundance of gaseous potassium in the former case of ``no rainout.''
The condensation cloud formation model predicts a higher abundance of
gaseous potassium in the upper part of the atmospheres of T dwarfs
(compare Figs. 2 and 3 of Burrows et al. 2000).  In Fig. 4, we compare
the spectrum computed with the K I abundance from our calculation without
``rainout'' (solid curve, Burrows \& Sharp 1999)  and with the
condensation cloud formation model (dotted curve, Lodders 1999).  The
latter agrees much better with the observed spectrum in the region
dominated by the K I resonance doublet. This further strengthens the case
for the formation of condensate clouds in the atmospheres of T dwarfs as
argued by Fegley \& Lodders (1996) and Lodders (1999).

The K I doublet at 1.2432 and 1.2522$\,\mu$m is weak but detected in our
spectrum, while the model predicts very strong lines.  Decreasing $\Teff$,
increasing $g$, and decreasing the potassium abundance could all reduce
the strength of the lines in the model.  The magnitude of the changes
required in these parameters is unreasonable in view of the other
constraints, however.  The two lines of the doublet share the same lower
level of K I (about 1.6$\,$eV above the ground state) and are quite
sensitive to temperature.  In Gl~570D, this doublet is formed at the level
where $T \sim 1600\,$K.  The strength of the doublet decreases rapidly at
lower temperatures. The larger than observed strength of the computed K~I
doublet may be due to remaining uncertainties in the $(T,P)$ profile in
the atmosphere model (perhaps due to the effects of condensates deep in
the atmosphere). If this is the case, the 1.25$\,\mu$m doublet of K I
could become a powerful probe of the temperature profile in T dwarf
atmospheres.

A better fit to the spectrum of Gl~570D could be obtained with a more
detailed study.  While the deviations we find could be due to residual
uncertainties in the $(T,P)$ profile of the atmosphere or the presence of
a warm dust layer, compositional variations from a solar mixture of heavy
elements may play a role. Feltzing \& Gustafsson (1998) found that the
abundances of individual elements in Gl~570A are scattered around the
solar values by $\pm~0.4$~dex.  In particular, they find [O/H]=0.16,
[Na/H]=0.06, and [Ca/H]$=-0.11$ with uncertainties of $\pm 0.1 - 0.2$
dex. If the abundances of several other key elements (C, N, O, K, Si, and
Mg, for example) also deviate appreciably from their solar values, and if
we assume that Gl~570D shares the same atmospheric composition as the
primary star, the modeled spectrum would be altered. Such refinements go
beyond the scope of the present study.

\section{Summary} 

New and improved spectra and photometry of Gl~570D, one of the coolest
and least luminous brown dwarfs known, have been analyzed by spectral
modeling to yield more accurate estimates of luminosity, effective
temperature, and surface gravity. The observations demonstrate that the
increasing blueness at $J-K$ of T dwarfs with decreasing temperature
continues down to objects with effective temperatures of at least 800~K.
Comparison of model spectra with the observed spectra provides some
evidence for rainout of condensates and suggests that the K~I doublet at
1.25~$\mu$m may be a sensitive indicator of the temperature profile in T
dwarfs.

\acknowledgements

We thank A. Burrows for sharing his table of alkali metal abundances, R.
A. Donahue for a useful discussion, J.W. Liebert for an important
suggestion, and an anonymous referee for several helpful comments.  
T.G.'s research was supported by the Gemini Observatory, which is operated
by the Association of Universities for Research in Astronomy, Inc., under
a cooperative agreement with the NSF on behalf of the Gemini partnership :
the National Science Foundation (United States), the Particle Physics and
Astronomy Research Council (United Kingdom), the National Research Council
(Canada), the Australian Research Council (Australia), CNPq (Brazil) and
CONICET (Argentina).D.S. acknowledges support from NASA through grant
NAG5-4988. G.K. thanks NASA for support via grant NSG5-6734. Work by K.L.
was supported by NSF grant AST-0086487. Work by MM was supported by NSF
CAREER grant AST 96-24878 and NASA grant NAG5-8919. The United Kingdom
Infrared Telescope is operated by the Joint Astronomy Centre on behalf of
the U. K. Particle Physics and Astronomy Research Council.  We thank the
staff of UKIRT for its support of these observations and the UKIRT Service
Program for providing the L$^{\prime}$ photometry.

\clearpage

\centerline{*** FIGURE CAPTIONS ***}
\vskip 10pt

\figcaption {Spectrum of Gl~570D (this paper) and Gl~229B (Geballe
et al. 1996, Leggett et al. 1999), the latter divided by 2, with principal
molecular absorption features and the potassium 1.25~$\mu$m doublet
identified.  The Gl~570D spectrum is slightly smoothed and has resolutions
of 0.003$~\mu$m shortward of 1.4~$\mu$m and 0.006~$\mu$m longward of
1.4~$\mu$m; those of Gl~229B are 0.0015~$\mu$m and and 0.003~$\mu$m,
respectively.}

\figcaption{Evolution of solar metallicity brown dwarfs and giant planets
in effective temperature -- gravity space.  The heavy solid lines are
cooling tracks for objects with masses of 0.075, 0.07, 0.06, 0.05, 0.04,
0.03, 0.02, and 0.01$\,M_\odot$, from top to bottom, respectively (Burrows
et al. 1997).  Evolution proceeds from right to left; the isochrones
bracketting the age of Gl 570D are shown by dotted lines.  The band
crossing the center of the figure is the locus of all models with the
luminosity of Gl~570D model B. Filled symbols correspond to the models
listed in Table 3 and used for the present analysis.}


\figcaption{Comparison of synthetic and observed spectra for each of the
models shown in Fig. 2. The synthetic fluxes are not normalized to the
data but obtained from the radius of each model (Table~3) and the distance
of the Gl 570 system.  The molecules responsible for the spectral features
in each region are identified in the top panel.}

\figcaption{Two spectra computed from model B under different assumptions
for the computation of chemical equilibrium: that condensates do and do
not rain out of the atmosphere. In the former case the K~I abundance is
higher in the upper reaches of the atmosphere and a better fit to the
observed spectrum shortward of 1.1$\,\mu$m is obtained.}

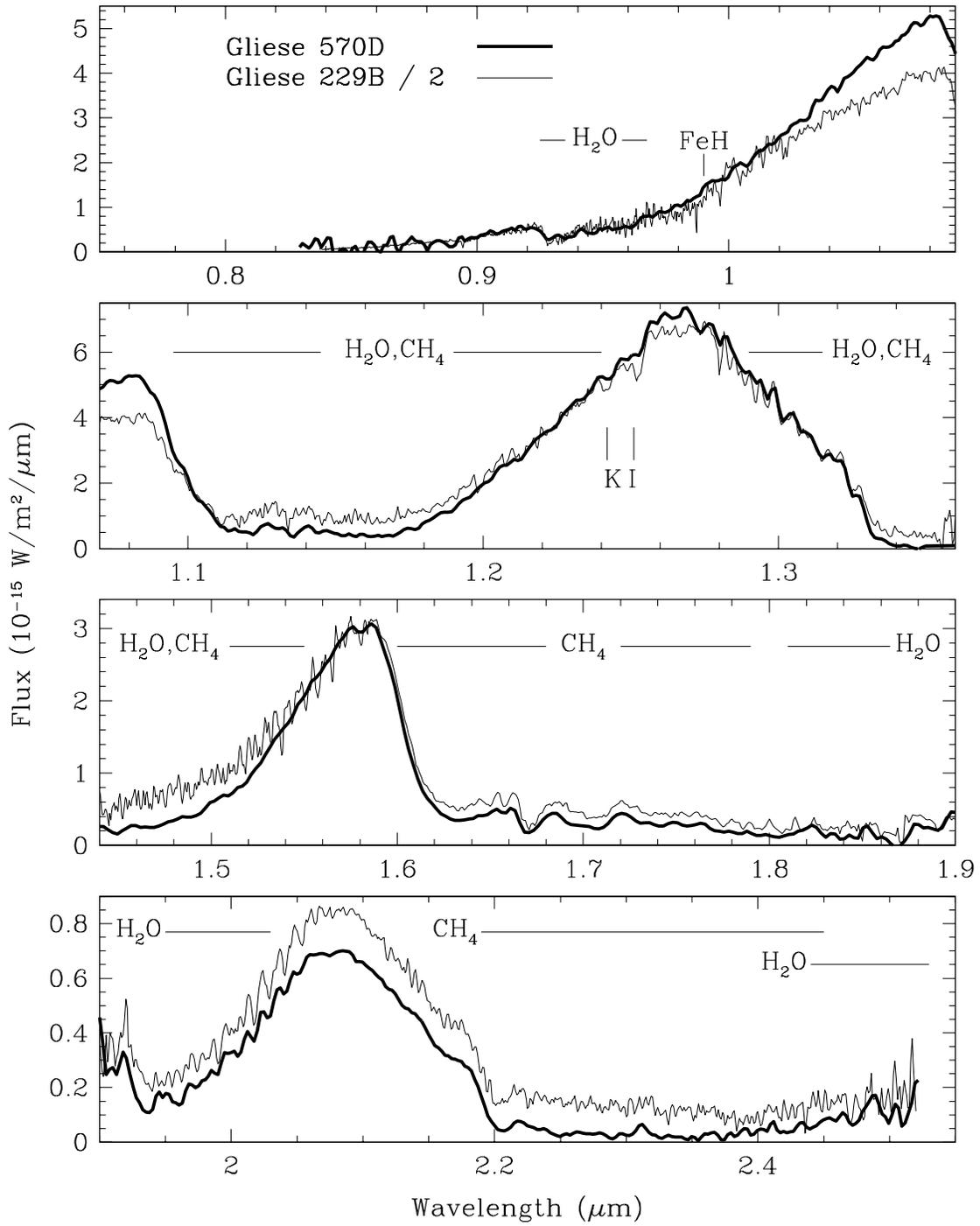
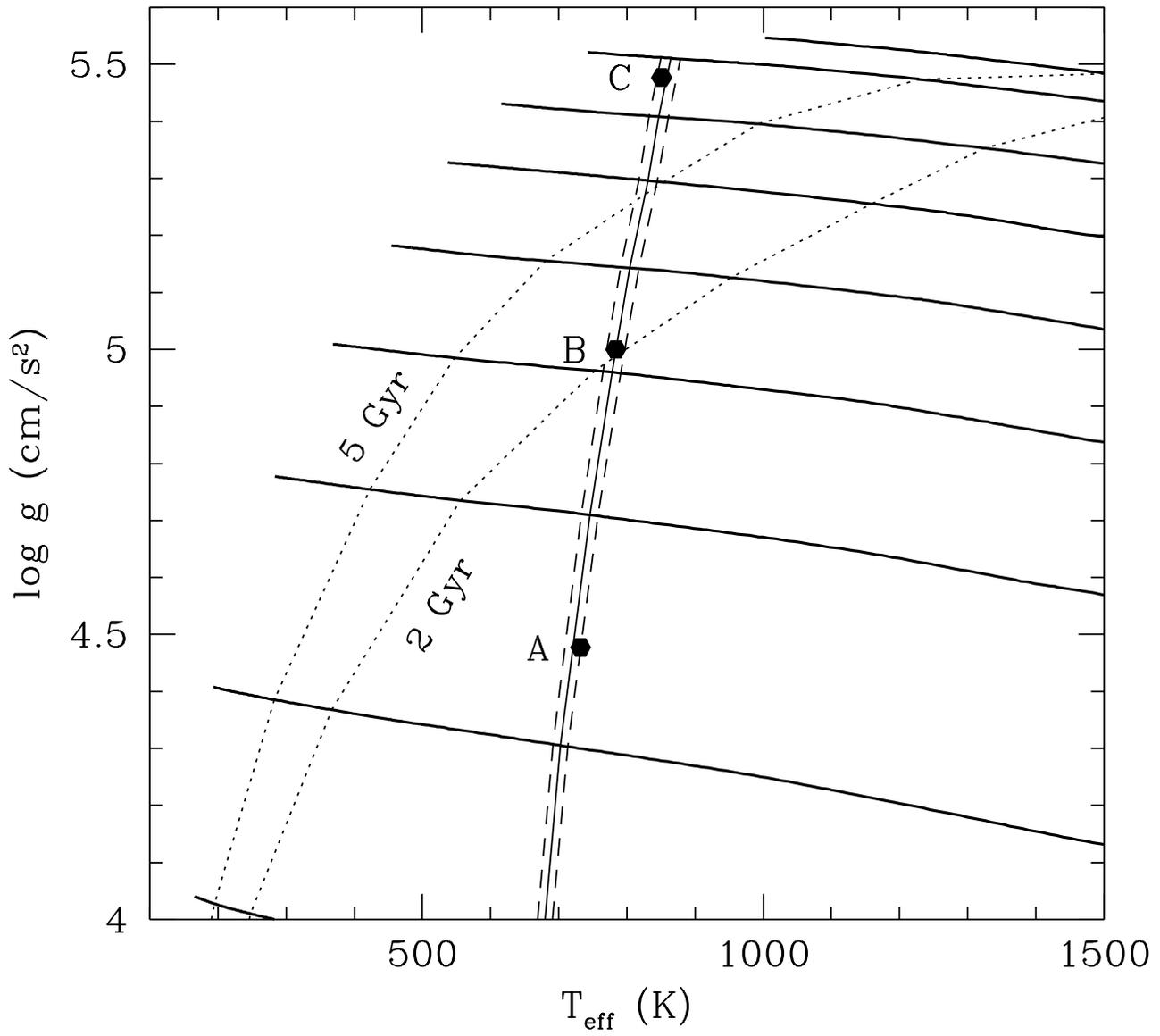
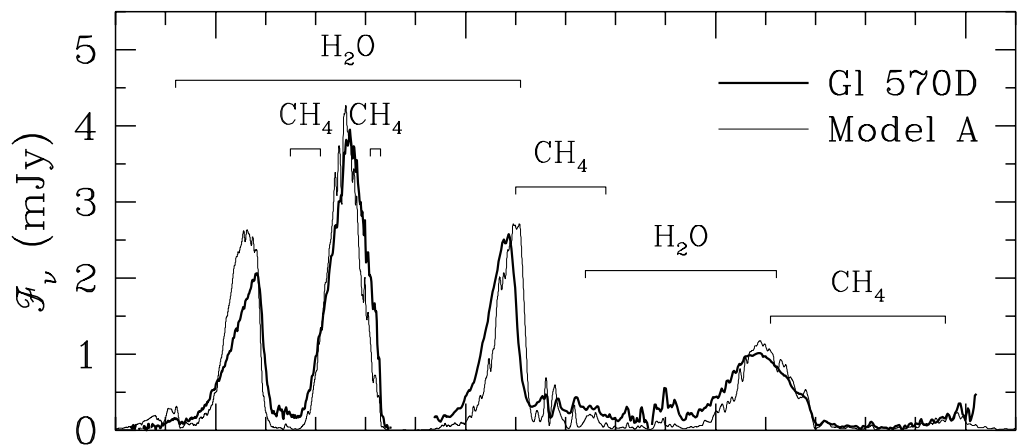
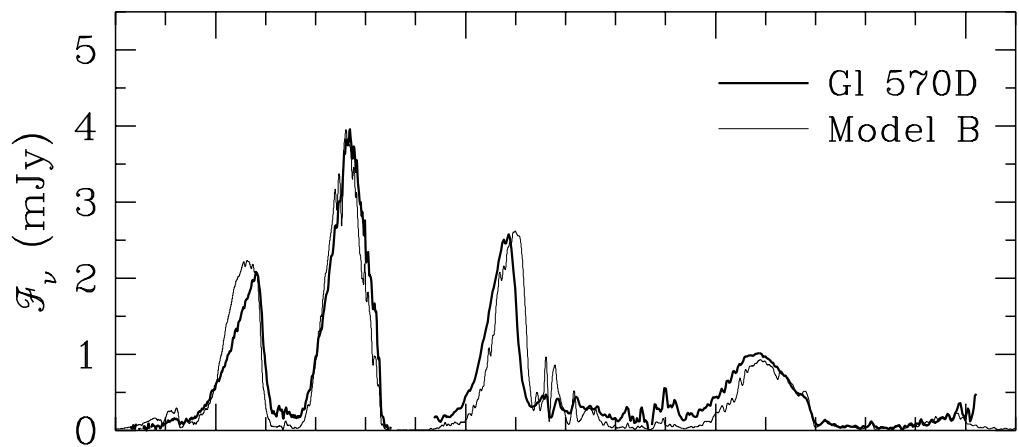
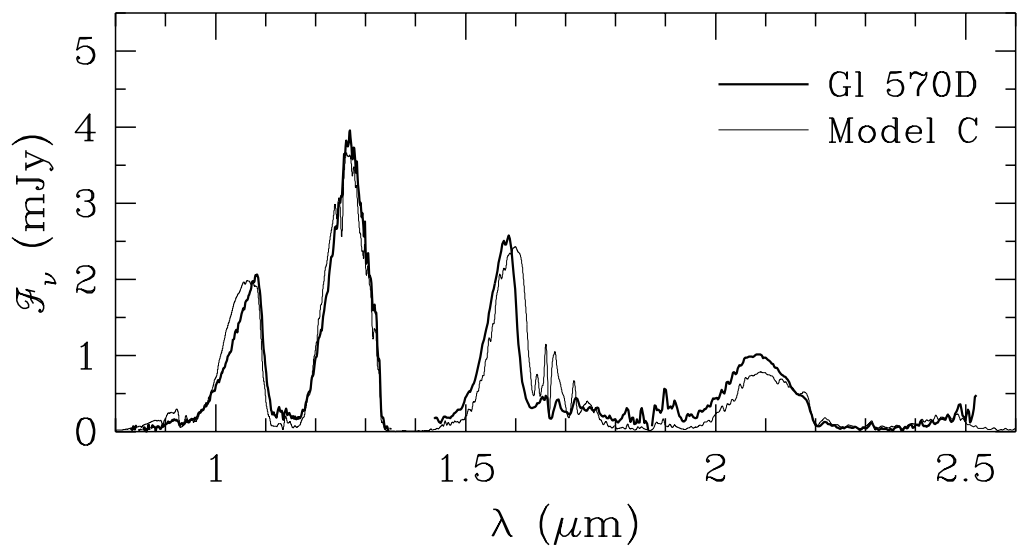
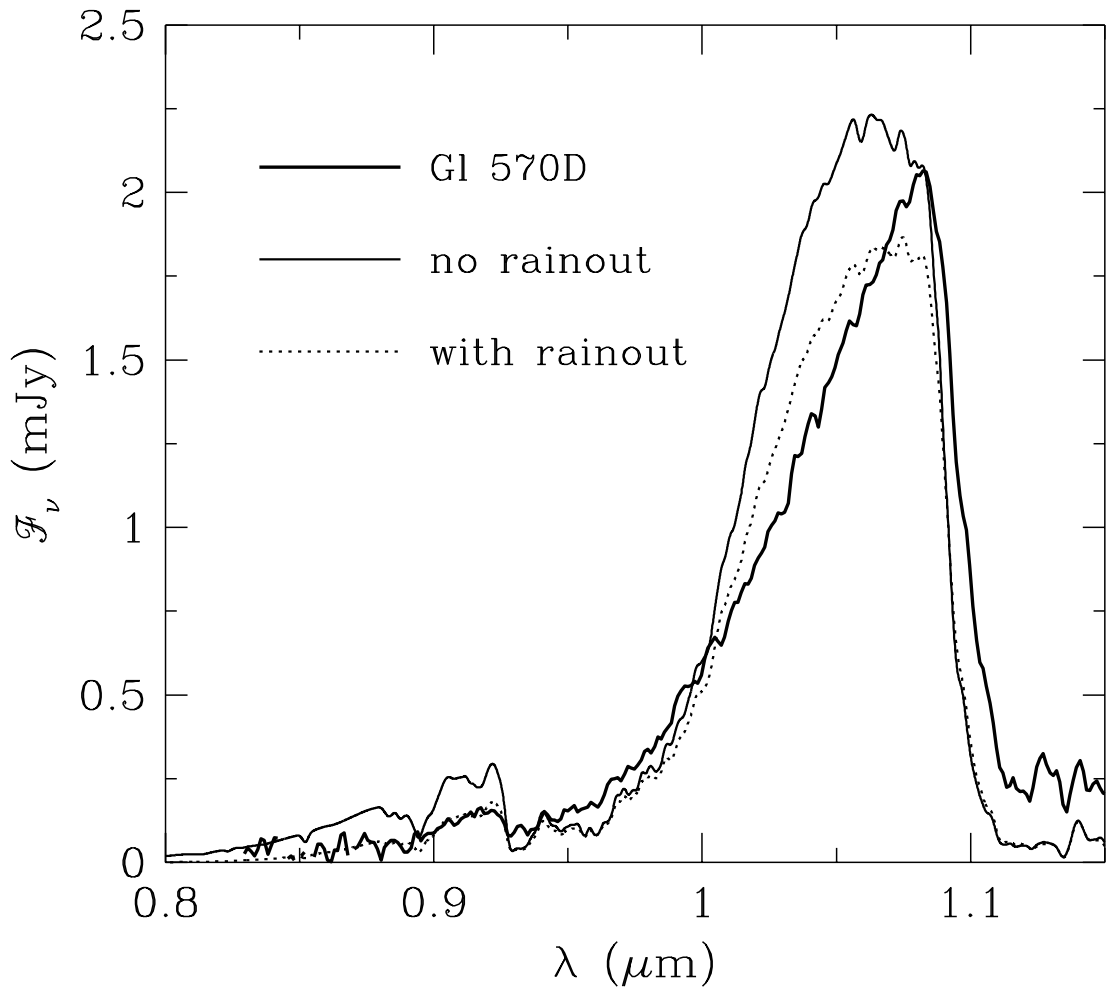
\end{document}